# Efficient Nano Antenna for Photonic Devices

Vishal K.Doltani
Institute of Information Communication Technology
M.U.E.T,Jamshoro
Sindh, Pakistan
vishal.doltani@outlook.com

Fahim A Umrani
Institute of Information Communication Technology,
M.U.E.T,Jamshoro
Sindh, Pakistan
faheemaziz.umrani@faculty.muet.edu.pk

Riaz A.Soomro
Institute of Information Communication Technology,
M.U.E.T,Jamshoro
Sindh, Pakistan
riaz.soomro@faculty.muet.edu.pk

*Abstract—* **This paper presents the efficient Yagi-Uda nanoantenna with the chain of directors, reflector and a feed element, where these elements have been optimized to fulfil the requirement of high directivity and the gain of antenna. The proposed design consists of six core and cladding with silver core and silicon cladding is used to achieve high directivity. The design is analyzed by FIT based CST Software package by which Directivity and Gain of the antenna are computed. The enhancement of this directivity is mainly contributed by the increase in the number of directors from the four to five and the optimized length of the directors which is scaled by 0.9. The simulated results show that the proposed Yagi antenna provides good performance in terms of directivity. The suggested antenna design shows the improved directivity of 17.62 at 500 nm wavelength. The proposed design has wide range of applications including as solar cells.**

I. INTRODUCTION

A device used for transmission and receiving electromagnetic waves is known as antenna [1]. The conversion of guided waves into free space waves is due to antenna. In recent years, many small-scale antennas at optical frequencies are developed. There are many types of conventional antennas available; for example, horn, bow-tie, dipole and many other antennas [2]. Nano antennas are widely being used by modern researchers for optical applications to improve the efficiency of light matter interaction using Nano photonics [3-6]. Different antennas have been used in photonics and one of them is a Yagi antenna which is well known in the low frequency radio systems [2] for its directional behaviour.

Yagi-uda antenna design was given by a young engineer, Shintaro Uda, Assistant Professor of Hidetsugu Yagi from Tokyo imperial university 1909 under the supervision of German Professor Heinrichs Barkhauson. Recently the traditional Yagi-uda antenna has produced light emission & detection that is unidirectional at nano-scale [3]. In [4], yagi antenna is designed using parasitic elements of reflector, directors and the feed element for achieving constructive interference in one direction and the destructive interference in opposite direction. In previous years many studies focused on the near field enhancement but lately the directionality of far field nanoantenna attracted more attention [5].

The paper is organized as follows. Section 2 contains literature review of Yagi uda Nano antenna and its properties. Section 3 gives the structure of Yagi antenna. Section 4 shows the fabrication types of Yagi antenna design. Simulation and methodology is defined in section 5. Section 6 shows the maximum directivity of 17.64 achieved .Section 7 conclusion of paper and Section 8 gives reference to future work which could be done using CST Simulation software.

II. LITERATURE REVIEW

Taminiau et al 2008, the Yagi-uda antenna was given the maximum directivity of 6.4 in contrast the dipole classical RF design antenna **[9]**. Liu et al 2012, Selective photonic sphere array has been used and the directivity of 15.7 at the wavelength of 603nm wavelength was achieved **[13]**. Krasnok et al 2012, due to the magnetic resonance in the visible range optical nanoantenna was made of silicon nanoparticles and the directivity of 12 at 500nm had been achieved **[14]**. Satitchantrakul and Silapunt 2014, Double driven element created for Yagi –Uda antenna in this paper, it is proposed that as the no: of directors is increased the directivity is improved but lastly the distance between double driven elements and first director decreased the directivity of 12.54 found at 660nm [15]. Huang et al 2016, the row of nanorods have been used at 400nm and increased 40% directivity [11]. Ma et al 2016, Substrate of SiO2 was used for the design of Yagi-Uda antenna at the wavelength of 633nm and the approximate directivity reached at 17 **[16].** Ghanim et al 2016, Yagi-Uda nano antenna was designed by using different materials in core and cladding with silver core surrounded by a silicon shell at the wavelength of 500nm which has given the approximate enhancement of 43.41% and directivity of 17.21 was achieved [8]. Yagi antenna properties shows optical Huygens source. which consists of a dipole electric operates at the resonance of a nanosphere. Such a structure shows high directivity and backward scattering disappears, construct attractive design for efficient and compact optical nanoantenna's[14].

### III. Yagi Antenna Structure

It is the chain of active and passive elements. It consists of one active and some passive elements [5] also known as driver and parasitic elements [6]. The active element is known as dipole that works in quarter one wavelength resonance and its structure is simple, low cost and easy to build and provide required results for many applications [6]. The passive element includes one reflector and one or more than one director as shown in **Fig. 01**. The active element includes one feed element known as dipole with the excitation gap [7]. The reflector and directors are designed to radiate in forward direction without any backward radiation. So that we consider longer length of reflector than feed element [8]. The energy flow of antenna would be in forward direction if it is properly tuned and energy would be distributed from dipole to the directors. The length of feed element is the key parameter for whole structure and it should be in resonance at desired frequency [3]. Additionally, Yagi-Uda Nano-antennas can lead to high directivity than other Nano antennas [9] such as a plasmonic nanowire [10], antenna structure like honeycomb [11] and a dipole placed near ring reflector [12].

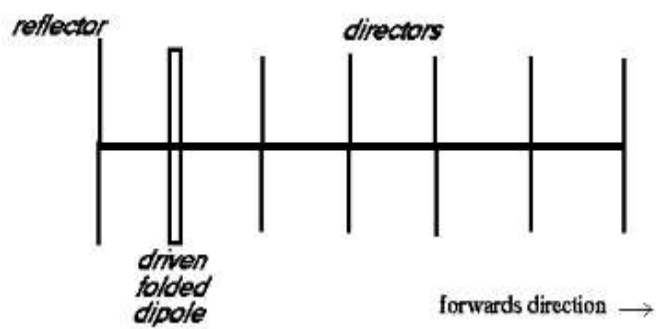

*Fig. 01: Structure of Yagi-Uda Nano antenna*

### IV. Fabrication Of Yagi-Design

The systematic process of fabrication of Yagi-uda nano antenna is performed using three different approaches. The fabrication of the suggested Au-Si nanoantenna can be performed by using the two-step electron-beam (EB) lithography evaporation technique [17]. Second method of fabrication is the vapour-liquid for the synthesis core-cladding nano particles for coaxial hybrid nanowires [18]. Third method for fabrication involves multiple steps. Initially nano antenna is directly fabricated by focused ion beam and refined by two photon photoluminescence, after that reactive ion is used for printing process [19] then finally a indexing match cap layer used for nanoantenna and it is excited by Nano emitter e.g. quantum dot or atom. Another recommended fabrication method is femto second laser used for multilayer thin film known as ablation method in which material is removed via laser pulse which creates free atoms and clusters, and this creates magnetic optical response and plasmonic excitation in visible range [20]. Another way to fabricate nanoantenna is lithography electron beam on transparent glass substrate. The system states the structured in spin coated film [21] via metal evaporation and a lift-off procedure [22].

TABLE. 1 Yagi-Uda Nano-antenna Parameter's

| Yagi-Uda Nano antenna Parameter's | | |
|---|---|---|
| Reflector radius | $R_{core}$=75nm | $R_{cladding}$=45 nm |
| Director Radius | $R_{core}$=70nm | $R_{cladding}$=42 nm |
| Length of reflector $L_r$ | 220 nm | |
| Length of feed $L_f$ | 160 nm | |
| Length of director $L_{d_1}$ | 144 nm | |
| Length of director $L_{d_2}$ | 129.6 nm | |
| Length of director $L_{d_3}$ | 116.64 nm | |
| Length of director $L_{d_4}$ | 104.976 nm | |
| Length of director $L_{d_5}$ | 94.478 nm | |
| Space b/w elements | 70 nm | |
| Scaling factor S | 0.9 nm | |

### V. Design Simulation & Methodology

The 3D Yagi-uda nanoantenna has been designed with silver (Palik) and silicon optical materials. It consists of 7 nanowires with silver (Palik) core and silicon cladding. Silver material has some best properties over other materials and its data can be taken from Palik data [23]. The length of reflector is considered as *Lr* and length of feed would be considered as *Lf* and the length of 1st director would be scaled by *0.9Lf and 2nd* director would be scaled by *0.9Ld1* and so on. The reflector radius of Silver Core is R1 and Silicon Cladding is R2. The radius of each director is R3 and R4 respectively. The radius data has been collected from golden ratios that is the ratio of 1.618 [23] between the radius of core and cladding. The space between each element is **70nm**. We are considering silver (Palik) data due to its high refractive index of 1.5 whereas the Johnson and Christy is 1.4 maximum [24]. The field element of nano antenna is excited by discrete port with the gap size of 10 nm between perfect conductor. The model has been designed through 3D numerical analysis technique known as Finite integration technique FIT [18] by using the computer simulation tool. Yagi-uda Nano antenna is defined in three coordinate systems x, y and z direction and open (add space) area boundary conditions surrounded by air as shown in **Fig. 02** and for achieving high accuracy,15 cells per wavelength have been used.

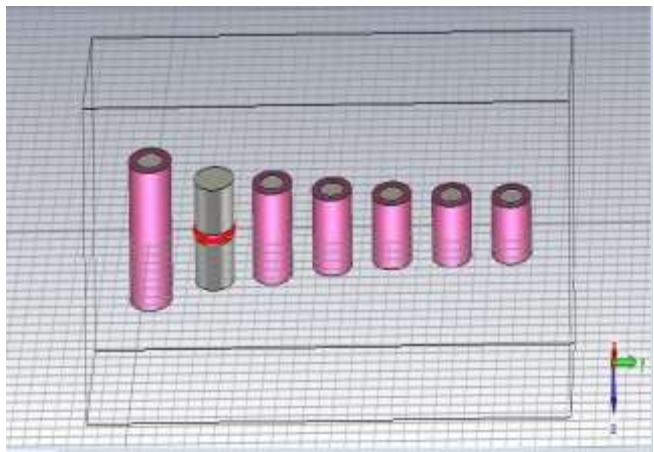

*Fig. 02: Yagi uda nano antenna design*

We consider geometric parameters of directivity from the Lorentz 4 parameters [5]. The numerical model for directivity is shown in **Eq.01**.

$$D = \left( 2.549 - \frac{8.546}{1 + \left(\frac{\frac{r_d}{r} - 0.805}{0.173}\right)^2} \right) - a, \qquad (01)$$

Where,

D shows the directivity of nanoantenna

The efficiency of nanoantenna can be calculated by radiated power divided by radiated power and the total power loss as shown in the **Eq. 02**.

$$\eta = \frac{P_{rad}}{P_{rad} + P_{loss}} \qquad (02)$$

VI. RESULTS & DISCUSSIONS

The proposed designed has been optimized by considering the Krasnok et al. [14], performance of nanoantenna can be predicted when the radius of director match to the magnetic resonance and electric resonance of reflector; through the coupling between elements at given frequency. The proposed Nano antenna consists of reflector with radius of reflector Rr = 75 nm and director with the radius of 70 nm and the separation between reflector and director is 70 nm. We find out results of efficient Yagi-Uda nanoantenna by plotting directivity versus wavelength and it has been calculated by using FIT technique. We noticed from the figure that good results are obtained in contrast with Krasnok et al design. We achieve maximum directivity of 17.62 at 500 nm and the increase of 47 % is achieved and the outcome of the material nature of the proposed design is indicated as shown in **Fig. 04.**

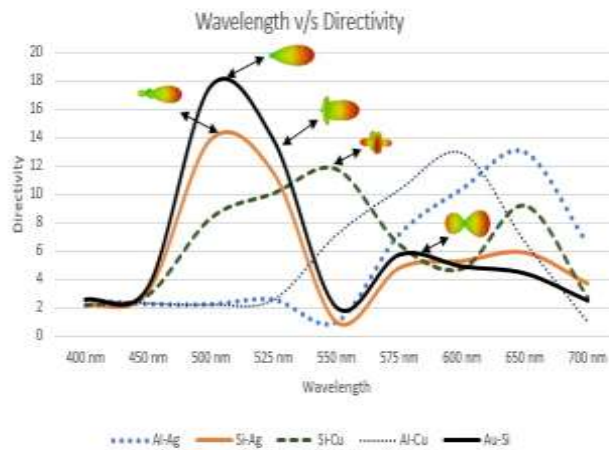

*Fig. 03: directivities of nanoantenna design using different core and cladding materials.*

Fig. 03 shows the outcome of the proposed design using different material types such as gold and aluminium for the core cladding nanowire nanoantenna. It can be seen in the figure that the directivity reaches about ∼12 for the copper–silicon design at 550 nm while the gold core increases the directivity to ∼13 at 600 nm. However, high directivity of ∼17.64 can be achieved by using silver core with silicon (Palik) cladding. It is revealed from Fig. 03 that the Si-Ag shell offers directivity of∼14, while the Silica shell has directivity of∼7 at a wavelength of 700 nm. Figure 03 illustrates the performance of five different materials combination with different wavelength. It is evident that the highest directivity gain of 17.62 is achieved using Au-Si at 500 nm.

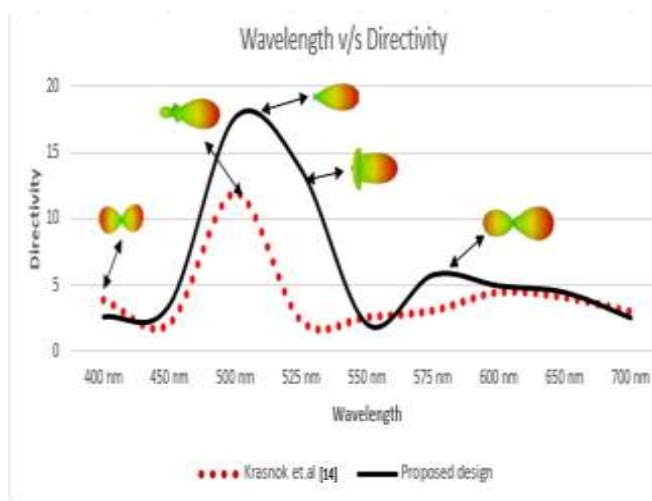

*Fig. 04: Wavelength dependent directivities of the proposed nanoantenna design with the spherical krasnok design [14] and radiation patterns of proposed design and spherical nanoantenna [14] at wavelength of 500 nm.*

The far field plot of antenna shows the radiation pattern of the antenna and it is defined in spatial coordinates that is defined by the azimuth angle (φ) and the elevation angle (θ). More precisely the plot of the power radiated from an antenna per unit solid angle as shown in **Eq. 03**. It has been plotted in 3D design as shown in **Fig. 05**. It is an extremely useful parameter as it shows the gain as well the directivity of antenna at various points in space. The directivity of nano antenna is defined in terms of azimuth and elevation as given in the equation **Eq. 03.**

$$D(\theta, \Phi) = 4\pi \frac{P(\theta, \Phi)}{\int P(\theta, \Phi) d\Omega} \qquad (03)$$

Where,

$P(\theta, \Phi)$ is known as the power radiated per unit solid angle.

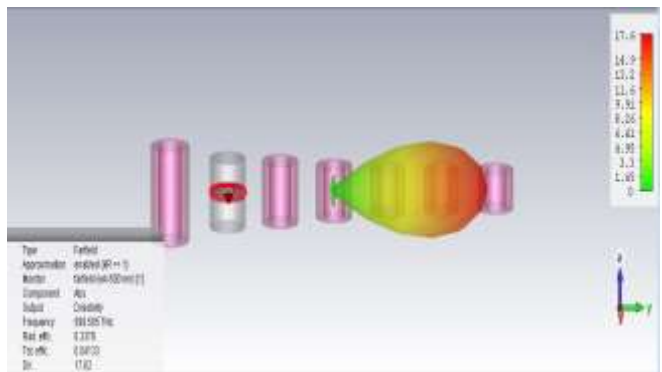

*Fig. 05: The directivity in the forward direction (in the y direction) of the photonic Nanoantenna as a Wavelength function*

We have noticed from **Fig. 06** which shows the TE field of the proposed design at the wavelength of 500nm that the enhancement in directivity is due to the uniform field distribution of electric and magnetic field distribution. On the other hand, in previous designs the Electric fields were non-uniform.

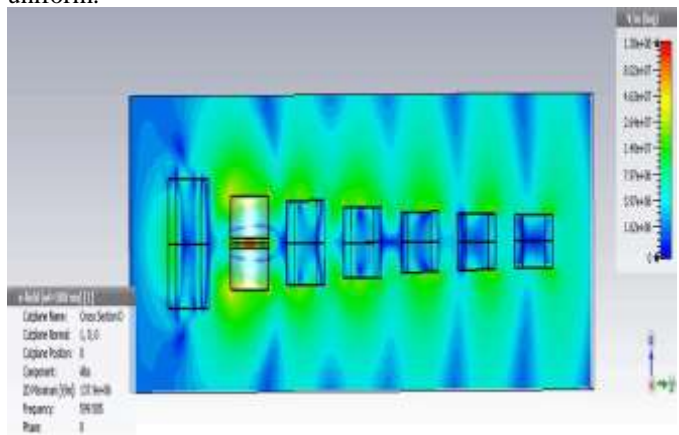

*Fig. 06: Uniform TE wave distribution over the field.*

Fig: 07 (a) shows the angular distribution of the radiated

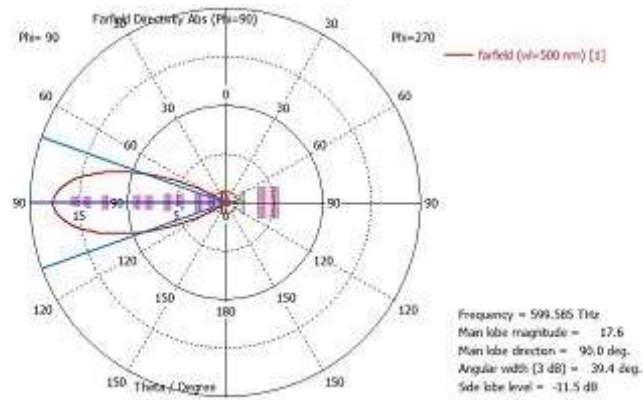

pattern (phi, Theta). Fig. 05 shows the directivity of proposed design in one direction at the wavelength of 500nm. Further the directivity of the cylindrical nanoantenna shows the replication of angular distribution of radiated pattern at λ = 575 nm is shown in Fig 7(b).

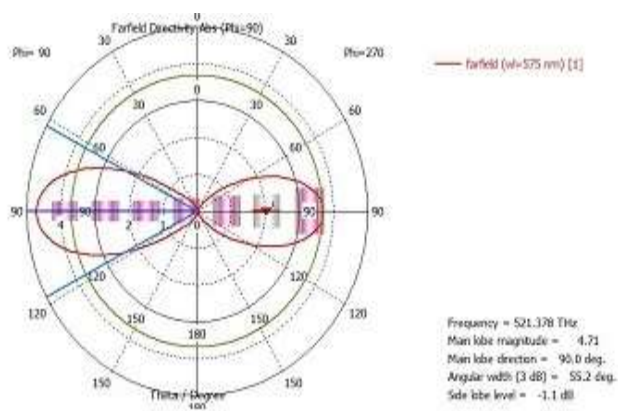

*Fig. 07(a) The radiation pattern at the (Θ, φ) plane after altering the parameters of yagi antenna at the wavelength of 500nm shows the maximum main lobe with directivity of 17.62*

*Fig. 07(b) angular radiation pattern with maximum back lobe at the wavelength of 575 nm*

Fig. 07 (a) and (b) compares the directivity caused by difference in wavelength with non-cooperative approach, where radiation flows common direction except in other directions respectively. The graph is plotted in theta and phi.

It may be observed that the radiation pattern of the proposed nanoantenna is more directive at the considered wavelength than the spherical nanoantenna [14]. The first sight on the graphs reveals that it has single main lobe in particular direction with the maximum directivity of 17.62. Whereas Fig. 7 (b) shows the back lobe with the minimum directivity of 4.71.

The performance of proposed optical nanoantenna design has been improved significantly by using high permittivity materials and this feature shows low loss at optical frequency. In our study we concentrate on nanoparticles prepared of silicon and silver (Palik) material at the 500nm wavelength that lies in the blue band spectrum and it is best material used for quantum dot effects which is currently in used for the optical communication technology.

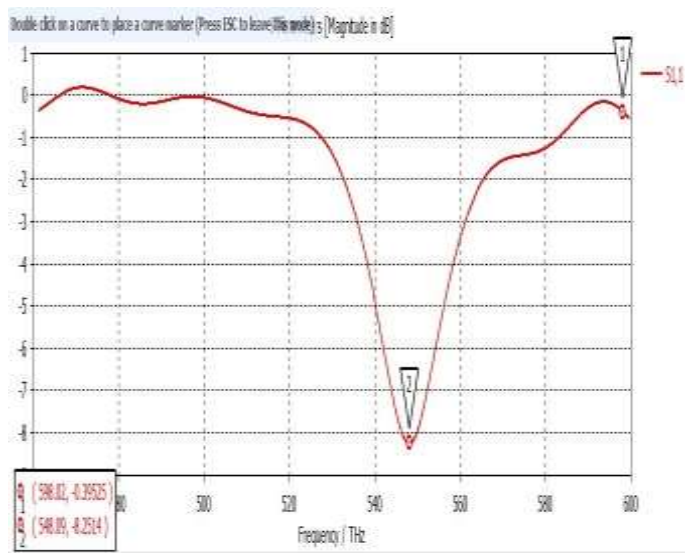

*Fig. 08: S-parameter chart for the efficient nanoantenna*

Fig:08 shows the S1,1-parameter return loss v/s frequency curve for the proposed antenna. where a better impedance matching and return loss is fine-tuned at 550 THz (-8.2 dB).

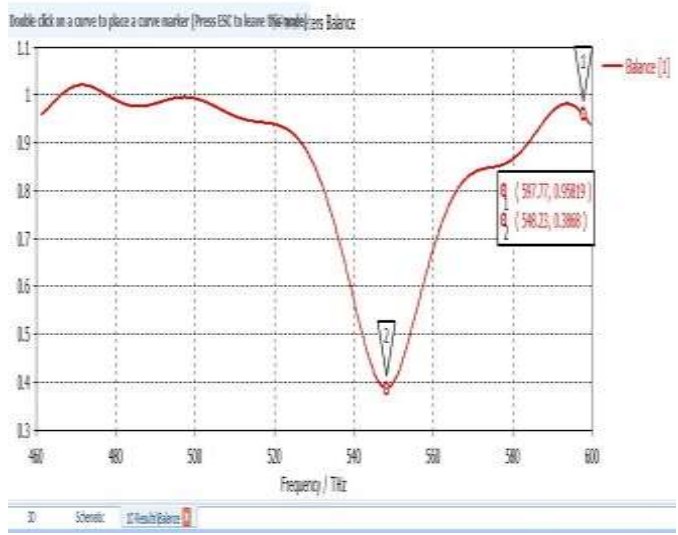

*Fig:09. Power loss v/s Frequency balance*

Fig. 09 shows the Balance (that is power loss versus frequency) of close to one which shows that the proposed structure is close to loss-less at wavelength of interest.

### VII. CONCLUSION

From above discussion we concluded that Yagi-uda Nano antenna is widely used due to its forward gain capability, easy construction and low cost. The Proposed model of Yagi-Uda Nano antenna has been analysed and improved from 12 to 17.62. So, the enhancement of this directivity was mainly added by the increase in the number of directors from the earlier four to five while considering the length of these directors together with the length of the reflector element. In proposed design of 6 element Yagi antenna, the length of the directors is scaled by 0.9. Hence, the use of 6 element in the proposed yagi antenna with carefully design will improve the quality of coverage due to increased directivity and henceforth the gain in the reception side and for the digital photonics devices has been achieved. This will lead to enhanced the quality of suggested Yagi-Uda nanoantenna.

### VIII. FUTURE WORK

Yagi antenna has been designed to convert photovoltaic energy from spectrum of EM waves into optical signals. Yagi antenna can be more improved by changing material and its parameters. In future its influence will be in numerous fields like broadband wireless links, mobile communication (5G), space communication, wireless optical communication and higher order frequency applications.

### ACKNOWLEDGMENT

This work is supported by Institute of Information & Communication Technologies, Mehran University of Engineering & Technology Jamshoro, Pakistan, for a M.E project under FEECE. The Authors are grateful to Department Telecommunication Engineering for facilitating lab equipment. Special Thanks to Engr. Riaz Ahmed Soomro, for his cooperation and assistance.